\documentstyle[11pt,IAU207_pasp,twoside,psfig]{article}
\def\edcomment#1{\iffalse\marginpar{\raggedright\sl#1\/}\else\relax\fi}
\reversemarginpar

\begin{document}
\title{Globular Cluster Systems in the Hydra~I Galaxy Cluster}
 \author{Michael Hilker}
\affil{Departamento de Astronom\'\i a y Astrof\'\i sica, P.~Universidad 
Cat\'olica, Casilla 104, Santiago 22, Chile}

\begin{abstract}
In this contribution, first results of deep VLT photometry (V,I) in the
central region of the Hydra~I galaxy cluster are presented.
Many star clusters have been identified not only around several early-type 
galaxies, but also in the intra-cluster field, as far as 250 kpc from the
cluster center. Outside the bulges of the
central galaxies NGC~3311 and NGC~3309, the intra-cluster globular cluster
system is dominated by blue clusters whose spatial distribution is similar
to that of the (newly discovered) dwarf galaxies in Hydra~I. 
The color distributions of globular clusters around NGC~3311 and NGC~3309
are multimodal, with a sharp blue peak and a slightly broader distribution of 
the red cluster population. 
\end{abstract}

\section{Introduction}

The study of globular clusters (GCs) in the center of galaxy clusters 
always has been of special interest due to their extraordinary rich abundance 
around the central (often cD) galaxy. Whether all central GCs
belong to the bulge of the central galaxy or to the cluster as a
whole is under discussion. Kinematic studies of GCs around NGC 1399 in the 
Fornax cluster (Richtler et al., this volume; Kissler-Patig et al. 1999) and 
M87 in the Virgo cluster (Bridges et al., this volume; Cohen \& Ryzhov 1997) 
seem to indicate that most GCs
outside a certain radius follow the gravitational potential of the
galaxy cluster rather than that of the galaxy. Also the detection of GCs far 
outside the center of NGC 1399 (Dirsch et al., this volume) seems to confirm 
the existence of intra-cluster globular clusters, as predicted by some authors 
(e.g. West et al. 1995, and references therein).
However, the process that is responsible for the formation of a very rich 
central GCS as well as a cD halo has not been understood yet.
The main scenarios are {\it i)} a multi-phase {\it in situ} formation within a 
common host halo (Forbes et al. 1997), {\it ii)} stripping of GCs from nearby 
galaxies (Kissler-Patig et al. 1999), or {\it iii)} formation in small 
sub-halos and the subsequent 
assembly into a giant elliptical, as for example the accretion of a large 
number of dwarf galaxies (e.g. C\^ot\'e et al. 1998; Hilker et al. 1999).
In order to decide which one of the processes is 
dominant, further observational constraints are needed.

The availability of 8m-class telescopes has made it possible for the first
time to investigate globular cluster systems in cluster of galaxies in great
detail beyond the nearby Virgo and Fornax cluster. At the distance of the 
Hydra~I cluster ($\simeq44$ Mpc or $(m - M)_V \simeq 33.2$ mag; Struble \& 
Rood 1991), only a
few CCD fields are needed to cover a wide range in radius from
the cluster core to the outside. With good seeing conditions and moderate
exposure times, globular clusters can be easily 
detected down to the peak of their luminosity function ($V \simeq 26$).
Moreover, dwarf galaxies as faint as the Local Group dwarf spheroidals
($\mu_V \simeq 25$ mag/arcsec$^2$) can be resolved and morphologically 
classified, and their properties can be related to those of the globular 
clusters. The same is true for the extended stellar halo of the central cD
galaxies whose light can be followed out to faint surface brightnesses.
Such a dataset enables us to study not only the GCSs of individual galaxies,
but the GCS of the galaxy cluster as a whole.

\begin{figure}
\centerline{\hbox{
\psfig{figure=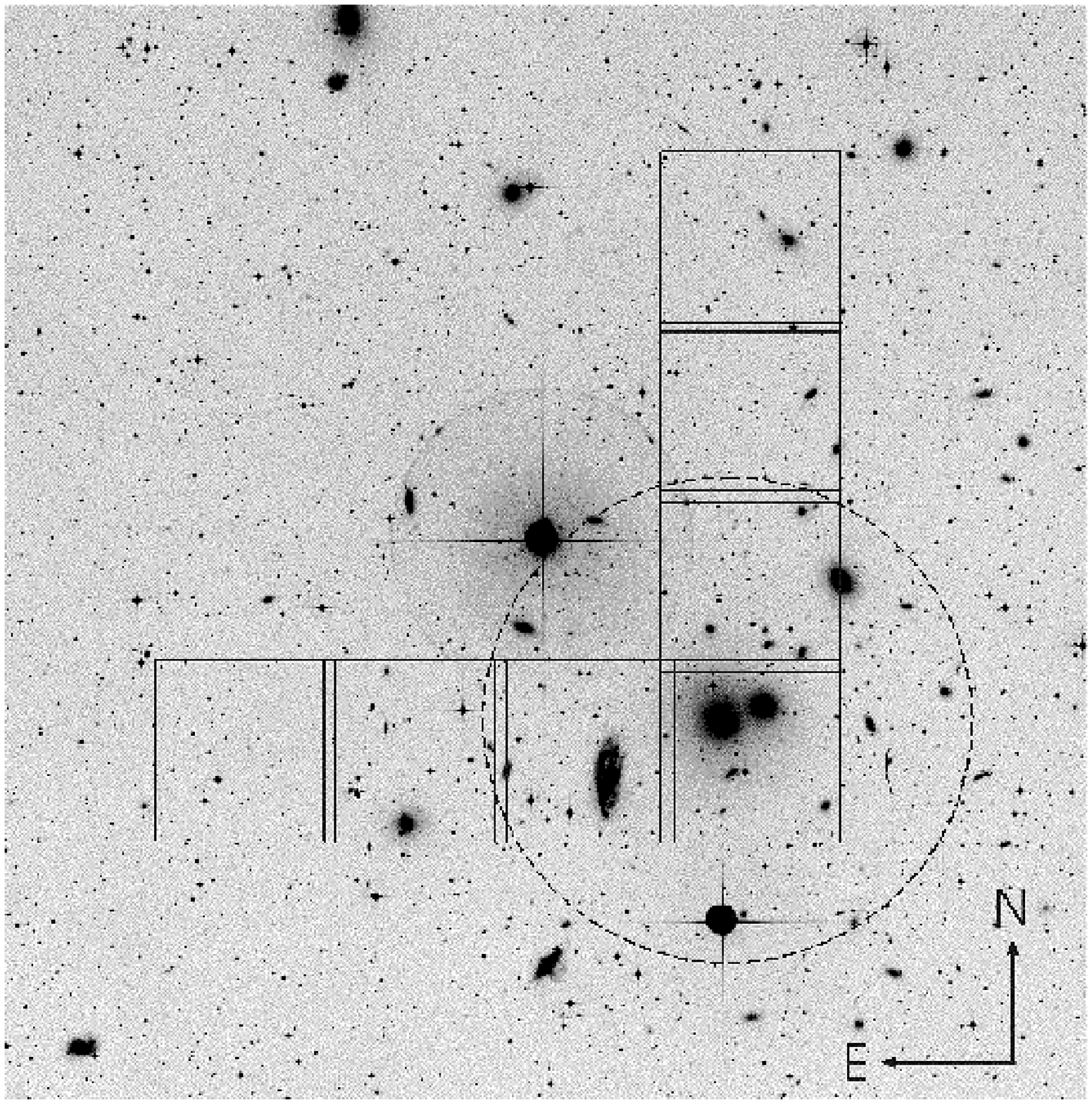,width=6.5cm,angle=0}
\psfig{figure=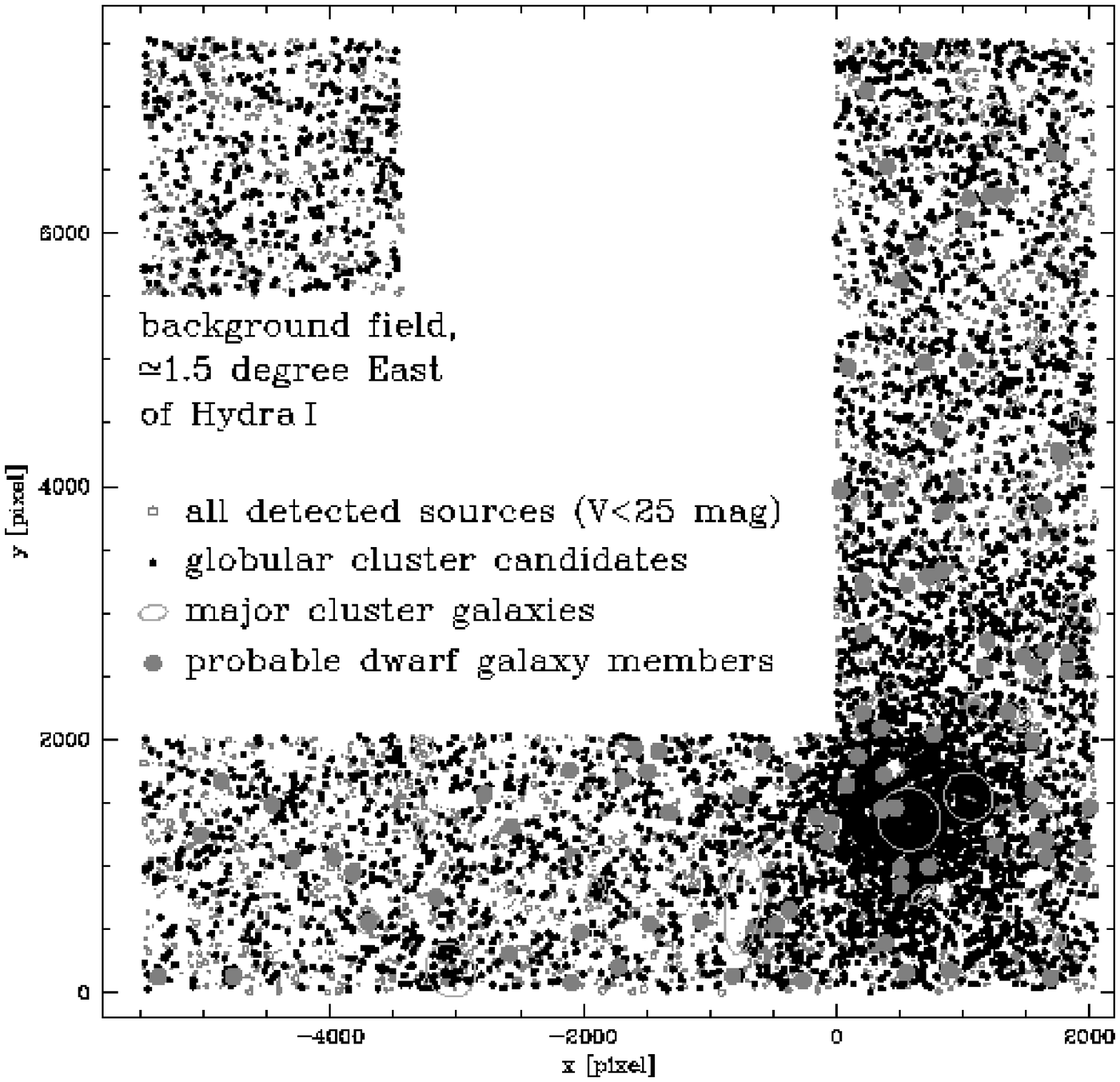,height=6.5cm,width=6.5cm
,bbllx=12mm,bblly=58mm,bburx=197mm,bbury=230mm,angle=0}
}}
\caption{On the left, a DSS image of the central region of the Hydra~I galaxy 
cluster is shown. The squares mark the position of the observed fields, the 
dotted circle the core radius of the cluster. In the right panel,
the distribution of all detected sources, globular cluster candidates, major 
galaxies, and probable dwarf galaxy members are shown.
Note the large extent of the central globular cluster system.
}
\end{figure}

In this contribution, I present first results on globular clusters in the 
Hydra~I galaxy cluster. This cluster has a compact regular core shape and a 
pair of bright galaxies in the center, whose brighter component, NGC 3311, 
possesses a cD halo and a extraordinary rich globular cluster system (e.g. 
McLaughlin et al. 1995). 

\begin{figure}
\psfig{figure=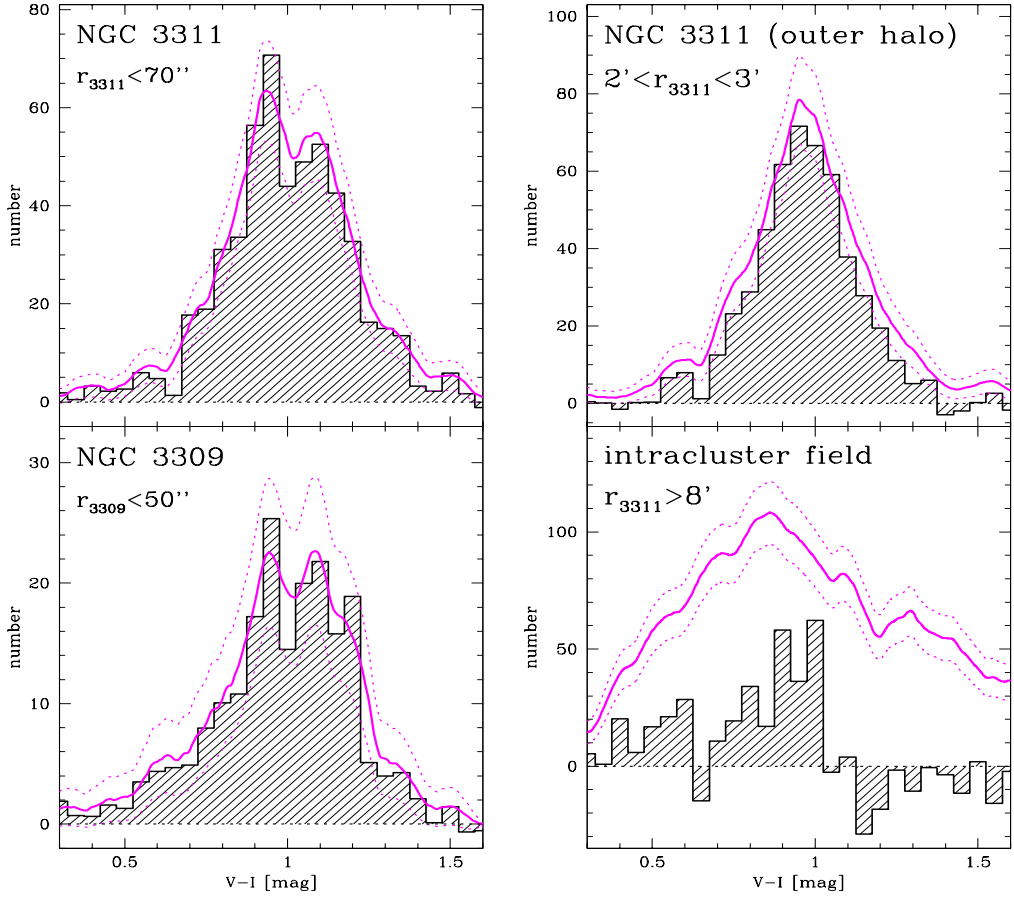,height=11.5cm,width=13.4cm
,bbllx=15mm,bblly=68mm,bburx=195mm,bbury=242mm,angle=0}
\caption{In this plot color distributions of globular cluster candidates
brighter than $V=25.5$ mag are shown. The hashed histograms are the background
corrected distributions. The solid curves show the uncorrected distributions
using a density estimation technique (Epanechnikov kernel). Whereas the 
histograms of NGC 3311 and NGC 3309 (left panels) indicate a clear separation 
between red and blue globular clusters, the globular clusters in the outer 
halo of NGC 3311 and the intra-cluster field are dominantly blue.}
\end{figure}

\section{Observation, reduction, and candidate selection}

Images of seven fields in the Hydra~I cluster (see L-shape configuration in 
Fig.1) and one background field have been taken through $V$ and $I$
filters. The observations have been performed in a service mode run in April
2000 with UT1/FORS1 at ESO/Paranal. All exposures were taken during dark time
and with a seeing between $0\farcs5$ and $0\farcs7$. 
The exposure times for $V$ and $I$ were $3\times8$ min and
$9\times5.5$ min, respectively.

On the combined images, the light of the main elliptical galaxies has been
subtracted by ellipse fitting before applying SExtractor for the finding and
photometry of the objects.
Point sources have been measured through an aperture of 2\arcsec diameter, and
then corrected for the total magnitude and photometric zero points from the ESO 
standard calibration plan.

All objects that have a FMHM$\leq$2\arcsec
and a color $0.4 < (V-I) < 1.6$ mag have been defined as globular
cluster candidates.
The adopted reddening throughout the analysis was $E_{V-I} = 0.10$ mag, the 
average of the values given by Schlegel et al. (1998) and Burstein \& Heiles
(1984).

\begin{figure}
\centerline{\vbox{
\psfig{figure=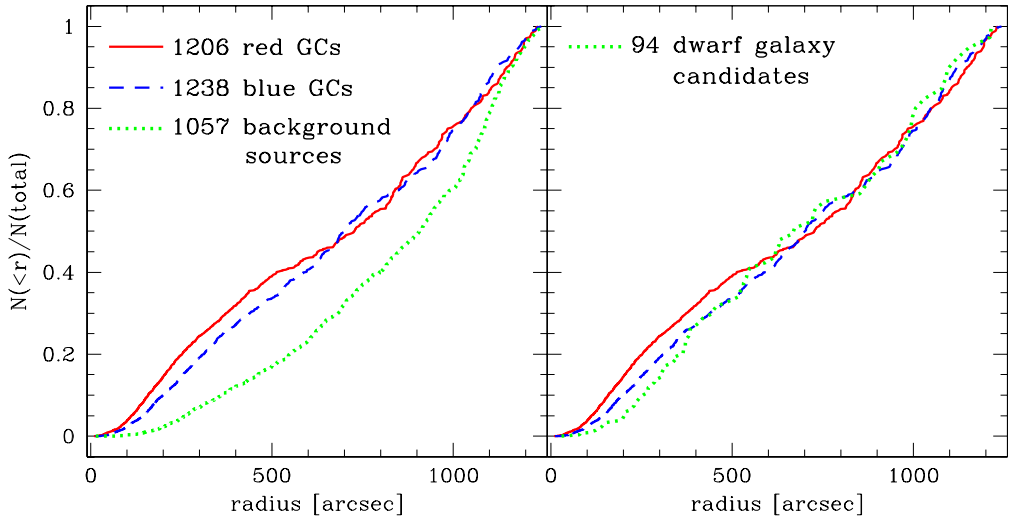,height=6.5cm,width=13.4cm
,bbllx=9mm,bblly=150mm,bburx=195mm,bbury=242mm,angle=0}
}}
\caption{This plot shows the cumulative radial distribution of red and blue
globular cluster sub-populations in Hydra~1. Additionally, the radial
distribution of background sources (left panel) and dwarf galaxy candidates
(right panel) are plotted. Within a 10\arcmin radius the red clusters are more
concentrated than the blue ones. The dwarf galaxies follow the distribution of
the blue cluster population.}
\end{figure}

\section{Globular Cluster color and spatial distribution}

Some years ago it was thought that there is a lack of metal-poor, 
blue GCs in central galaxy NGC~3311. Using Washington photometry, only red 
GCs with a mean metallicity of [Fe/H]=$-0.31$ dex were found (Secker et 
al. 1995). Only recently, this surprising result has been overruled by
HST photometry, which shows a normal color distribution of the GCS, comparable
to that of other giant ellipticals (Brodie el al. 2000). This has also been
confirmed by our analysis (see Fig.2). For the color distribution in Fig.2,
only point sources with $V<25.5$ mag and a photometric error less than 0.1 mag
have been selected. Applying the same selection criteria for the control field,
the color distribution of background sources has been subtracted statistically.
NGC~3311 as well as NGC~3309 possess
a blue globular cluster sub-population with a well defined peak at about
$(V-I) = 0.93$ mag ([Fe/H]$\simeq-1.5$ dex). The red sub-population has
a broader distribution peaking around $(V-I) = 1.08$ mag. Most probably,
this reflects a superposition of several red sub-populations of different
metallicities.
The existence of an extended blue tail in the color distributions
points either to a population of very metal-poor globular clusters 
([Fe/H]<$-2.0$ dex) or a population of young clusters of a rather recent 
star formation event (that might be related to the central dust lane in
NGC~3311).

In the outer halo of NGC 3311 (beyond 16 kpc from the galaxy center) blue
globular clusters start to dominate the distribution. The lower right
panel of Fig.2 shows that (mainly blue) globular clusters exist out
to very large distances from the cluster center ($<250$ kpc). In this plot
all globular cluster systems of the major galaxies in Hydra have been
excluded.

The dominance of blue globular clusters in the halo and the large extent
of the intra-cluster GCS is also reflected in the radial distribution
of the different sub-populations. Figure~3 shows the high concentration of
red clusters ($1.0<(V-I)<1.2$) within a radius of 10\arcmin. Interestingly,
the distribution of dwarf galaxy candidates (see Fig.1) closely follows the
distribution of blue globular clusters ($0.7<(V-I)<0.9$), except a possible
deficiency in the very center of the cluster. 

\section{Future Analysis}

The study of the color and spatial distribution has shown that the
investigation of globular clusters in galaxy clusters should not only be 
restricted to the analysis of GCSs around individual member galaxies, but all
the intra-cluster GCs has to be taken into account.
The next steps for further analyses of the presented data are:\\
\noindent
$\bullet$ Constructing a density and color map for the sub-populations of the
intra-cluster GCS in order to define its center and compare it to that of
the central galaxy and the X-ray gas halo.\\
$\bullet$ Modelling the cD halo light and study the specific frequency as a 
function of centro-cluster distance.\\
$\bullet$ Studying the individual GCSs for all member galaxies down to the 
dwarf galaxy regime (colors, density profiles, specific frequency, etc.).\\
$\bullet$ Confirming the membership of dwarf galaxies and bright compact objects
by follow-up spectroscopy (Magellan I data in hand).\\

\acknowledgements I wish to thank L. infante and F. Barrientos for their 
contributions in this project and FONDECyT for its support through `Proyecto 
FONDECYT 3980032'.

\section*{Discussion} 

\noindent {\it Grillmair:\, } {Has anyone ever found an offset between centers
of globular cluster systems and the centers of their host galaxies/galaxy
clusters?} \\

\noindent {\it Hilker:\, } {No, but our data set enables us to examine this for 
the first time. It would ber of special interest to investigate this issue in
clusters that show an X-ray halo whose center is displaced with respect to the 
position of the central galaxy.}\\

\noindent {\it Gallagher:\, } {Have you compared the radial distributions of
blue GCs in Hydra with that of the dEs?} \\

\noindent {\it Hilker:\, } {First analysis seems to indicate that the spatial
distribution of both populations are closely related (see Fig.3). However, the
membership of the individual dwarf galaxy candidates has to be confirmed
(by our follow-up spectroscopy for example) and their globular cluster systems
have to be identified, before further interpretations can be made.}


\begin{references}

\reference Brodie, J.P., Larsen, S.S., \& Kissler-Patig, M. 2000, ApJ, 543, L19
\reference Burstein, D., \& Heiles, C., 1984 ApJS, 54, 33
\reference Cohen, J.G., \& Ryzhov, A. 1997, ApJ, 486, 230
\reference C\^ot\'e, P., Marzke, R.O., \& West, M.J. 1998, ApJ, 501, 554
\reference Forbes, D.A., Brodie, J.P., \& Grillmair, C.J. 1997, AJ, 113, 1652
\reference Hilker, M., Infante, L., \& Richtler, T. 1999, A\&AS, 138, 55
\reference Kissler-Patig, M., Grillmair, C.J., Meylan, G., Brodie, J.P., Minniti, D., \& Goudfrooij, P. 1999, AJ, 117, 1206
\reference McLaughlin, D.E., Secker, J., Harris, W.E., \& Geisler, D. 1995, AJ, 109, 1033
\reference Schlegel, D.A., Finkbeiner, D.P., \& Davis, M. 1998, ApJ, 500, 525
\reference Secker, J., Geisler D., McLaughlin D.E., \& Harris W.E. 1995, AJ, 109, 1019
\reference Struble, M.F., \& Rood, H.J. 1991, ApJS, 77, 363
\reference West, M.J., C\^ot\'e, P., Jones, C., Forman, W., \& Marzke, R.O. 1995, ApJ, 453, L77

\end{references}
\end{document}